\documentclass[10pt,journal,twocolumn]{IEEEtran}
\usepackage{amsmath}
\usepackage{amssymb}  
\usepackage{cases}
\usepackage{graphicx} 
\usepackage{subfigure} 
\usepackage{epstopdf} 
\usepackage{indentfirst}
\usepackage{bm}
\usepackage{setspace}
\usepackage{caption} %
\usepackage{xcolor}
\usepackage{cite}
\usepackage{enumerate}
\usepackage{algorithm}
\usepackage{algpseudocode}

\newcommand{\wuhao}{\fontsize{18pt}{\baselineskip}\selectfont}

\newtheorem{lemma}{Lemma}

\begin{document}
\title{\wuhao Covariance-Based Spectrum Sensing for Noncircular Signal in\vspace{-2mm} \\ Cognitive Radio Networks With Uncalibrated Multiple Antennas}
\author{An-Zhi Chen and Zhi-Ping Shi, \IEEEmembership{Member, IEEE} \vspace{-0.4cm}
\thanks{A.-Z. Chen and Z.-P. Shi are with the National Key Lab of Science and Technology on Communications, University of Electronic Science and Technology of China, Chengdu 611731, China (e-mail: chen370951351@gmail.com; szp@uestc.edu.cn).}}
\maketitle
\vspace{-110mm}
\begin{abstract}
In this letter, the problem of spectrum sensing is addressed for noncircular (NC) signal in cognitive radio networks with uncalibrated multiple antennas. Specifically, by taking both the standard covariance and complementary covariance information of the NC signal into account, a new robust spectrum sensing method called NC covariance (NCC) is proposed, which can fully reap the statistical property of the NC signals. Meanwhile, we derive the asymptotic distribution of the NCC statistic under the signal-absence hypothesis and obtain the theoretical decision threshold of the NCC method. Simulation results demonstrate that the proposed method is capable of outperforming state-of-the-art methods.
\end{abstract}
\vspace{-1mm}
\IEEEpeerreviewmaketitle
\vspace{-2mm}
\section{Introduction} 
\IEEEPARstart{C}{ognitive} radio (CR) has been recognized as a powerful means to improve the spectrum efficiency by allowing secondary users (SUs) to opportunistically utilize the licensed spectrum of primary users (PUs) \cite{1}. One major functionality of CR is the spectrum sensing (SS), which plays an important role to sense idle spectrum bands and avoid unacceptable interference to PUs\cite{2}. In the past decade, various detection methods based on different operational requirements were proposed for SS\cite{3}. Among them, maximum-minimum eigenvalue (MME)\cite{4} is a very attracting method for SS with multiple antennas, due to its ability of overcoming the noise uncertainty and fading effects. However, the MME method is susceptible to unequal per-antenna noise variances as may appear in practical scenarios due to calibration errors.

In the literature, several robust multiantenna SS (MSS) methods have been proposed for overcoming this drawback, such as the covariance absolute value (CAV) \cite{5}, locally most powerful invariant test (LMPIT)\cite{6}, Hadamard (HDM) ratio test\cite{7}, volume-based detection (VD)\cite{8} and separating function estimation test (SFET)\cite{9} methods can deliver desirable performance under the scenario of unequal per-antenna noise variances. Nevertheless, all these methods above were devised for circular signals as they only use the standard covariance information. But practically noncircular (NC) signals, for instance offset quadrature phase shift keying (OQPSK), pulse amplitude modulation (PAM) and binary phase shift keying (BPSK) are often encountered in modern communication systems \cite{10}. These signals contain not only the standard covariance information but also the additional complementary covariance information, which can be used for enhancing the detection performance\cite{11}. Therefore, there has been a recent interest in developing new methods for MSS that exploit both the standard covariance and complementary covariance information. In this case, Huang \emph{et al.} \cite{12} proposed a NC-HDM method to exploit the NC characteristic of the primary signals for MSS, and they showed that the NC-HDM method can significantly improve the performance of the original HDM method. However, the calculation of the NC-HDM test statistic involves matrix determinant computation, which incurs significant computational cost. More recently, a NC local average variance (NC-LAV) method was developed in \cite{13}. However, the NC-LAV method relies heavily on the assumption that the noise variances at all antennas are identical, thereby its performance is also sensitive to unequal per-antenna noise variances.

In this letter, we consider the SS problem for NC signal in CR networks with multiple receive antennas, focusing on the scenario where the antennas experience different levels of noise power. To leverage the benefits of the NC signal, we propose a new robust SS method named NC covariance (NCC) for such networks. Specifically, we show that the standard covariance and complementary covariance matrices of the received signals differ between the null and the alternate hypotheses, which can be used for detecting the presence of PUs. Moreover, an asymptotic distribution of the NCC statistic under the null hypothesis is derived to acquire the theoretical formula for the decision threshold. In addition, we also analyze the computational complexity of the proposed NCC method. Simulation results show that NCC outperforms the existing state-of-the-art methods.

\emph{Notation:} Throughout this letter, we use boldface lowercase and uppercase letters to designate vectors and matrices, respectively. Superscripts $(\cdot)^\ast$, $(\cdot)^T$ and $(\cdot)^H$ indicate the complex conjugate, transpose, and Hermitian transpose, respectively. While $|\cdot|$ denotes the absolute value operation, $\mathbb{E}[x]$ is the expected value of the complex-valued random variable $x$ and $\mathbb{D}[x]= \mathbb{E}[|x|^2]- |\mathbb{E}[x]|^2$ denotes its variance. The symbols $\mathbf{I}_M$ and $\mathbf{0}_{M}$ refer to the $M \times M$ identity and all-zero matrices, respectively. The trace of a square matrix $\mathbf{X}$ is denoted by $\text{tr}(\mathbf{X})$. A random variable $a$ follows the real Gaussian distribution with mean $\mu_r$ and covariance $\sigma_r^2$ is shown by $a \sim \mathcal{N}(\mu_r, \sigma_r^2)$, while a complex-valued random variable $b$ follows the circularly symmetric complex Gaussian distribution (CSCG) with mean $\mu_c$ and covariance $\sigma_c^2$ is denoted by $b \sim \mathcal{N}(\mu_c, \sigma_c^2)$.
\section{Signal Model}
\subsection{Standard and Complementary Covariance Matrices}
Let $\mathbf{z}\in \mathbb{C}^{M\times 1}$ denote a zero-mean complex random vector. The matrix $\mathbf{R}_{\mathbf{z}}=\mathbb{E}\big[\mathbf{z}\mathbf{z}^H\big]$ is the standard covariance matrix of $\mathbf{z}$, which is a Hermitian positive-definite matrix. While the matrix $\mathbf{C}_{\mathbf{z}}=\mathbb{E}\big[\mathbf{z}\mathbf{z}^T\big]$ is the complementary covariance matrix of $\mathbf{z}$, and is complex symmetric. The random vector $\mathbf{z}$ is circular if its complementary covariance matrix $\mathbf{C}_{\mathbf{z}}=\mathbf{0}_{M}$ and noncircular if $\mathbf{C}_{\mathbf{z}}\neq\mathbf{0}_{M}$ \cite{10}. Therefore, to fully capture the complete statistical characterization of a zero-mean noncircular random signal $\mathbf{z}$, both $\mathbf{R}_{\mathbf{z}}$ and $\mathbf{C}_{\mathbf{z}}$ should be taken into account for the detector design.
\subsection{Problem Description} 
In this work, we consider the CR network consisting of one SU equipped with $M$ uncalibrated antennas and $q$ single-antenna PUs, where the SU attempts to detect the signals emitted by $q$ PUs. Denote $\mathcal{H}_{0}$ as the null hypothesis that all PUs are absent and $\mathcal{H}_{1}$ as the alternative hypothesis that at least one PU is present. Mathematically, the signal vector $\mathbf{y}(k)\in \mathbb{C}^{M\times 1}$ received at the SU corresponding to the $k$th time instant can be written as\cite{12}
\begin{equation} \label{Eq1}
    \begin{cases}
    \mathcal{H}_{0}:& \mathbf{y}_k = \mathbf{w}_k, \\
    \mathcal{H}_{1}:& \mathbf{y}_k = \mathbf{H}\mathbf{s}_k+\mathbf{w}_k,
    \end{cases} k=1,\ldots,K
\end{equation}
where $K$ denotes the total number of received samples, $\mathbf{w}_k\in \mathbb{C}^{M\times 1}$ is the zero-mean CSCG noise vector with an unknown diagonal covariance matrix $\mathbf{R}_{w}=\text{diag}\{\sigma^2_{1},\sigma^2_{2},\ldots,\sigma^2_{M}\}$, i.e., $\mathbf{w}_k \sim \mathcal{CN}(\mathbf{0}, \mathbf{R}_{w})$. Note that $\sigma^2_{m}$ is possibly unequal to $\sigma^2_{n}$ for $m\neq n$ due to the uncalibrated receiver in practice. $\mathbf{H}\in \mathbb{C}^{M\times q}$ denotes the channel matrix between the SU and PUs, which is assumed to be kept unchanged within a detection period. Additionally, $\mathbf{s}_k=[s_1(k),\ldots,s_q(k)]^T$ is the source signal vector, in which the symbol signal $s_i(k)~(i=1,\ldots,q)$ emitted by the $i$th PU is assumed to be noncircular (i.e., $\mathbb{E}[s_i^2(k)]\neq0$), identically distributed with mean zero and variance $\gamma_i=\mathbb{E}[|s_i(k)|^2]$. Meanwhile, we suppose that signal $\mathbf{s}_k$ and noise $\mathbf{w}_k$ are statistically independent.
\section{Proposed Noncircular Covariance Method}
Let us first consider the standard covariance and complementary covariance matrices of the received signal vector $\mathbf{y}_k$, given as
\begin{equation} \label{Eq2}
        \hspace{-3mm}\mathbf{R}_{\mathbf{y}}=\mathbb{E}\big[\mathbf{y}_k\mathbf{y}^H_k\big]~\text{and}~
        \mathbf{C}_{\mathbf{y}}=\mathbb{E}\big[\mathbf{y}_k\mathbf{y}^T_k\big].
\end{equation}
Under the null hypothesis $\mathcal{H}_{0}$, $\mathbf{R}_{\mathbf{y}}$ and $\mathbf{C}_{\mathbf{y}}$ are respectively given as
\begin{equation} \label{Eq3}
    \begin{aligned}
        &\mathbf{R}_{\mathbf{y}}|\mathcal{H}_{0}=\mathbf{R}_{w}=\mathbb{E}\big[\mathbf{w}_k\mathbf{w}^H_k\big]=\text{diag}\{\sigma^2_{1},\sigma^2_{2},\ldots,\sigma^2_{M}\}, \\
        &\mathbf{C}_{\mathbf{y}}|\mathcal{H}_{0}=\mathbb{E}\big[\mathbf{w}_k\mathbf{w}^T_k\big]=\mathbf{0}_{M}.
    \end{aligned}
\end{equation}
Meanwhile, under the alternative hypothesis $\mathcal{H}_{1}$, we can derive $\mathbf{R}_{\mathbf{y}}|\mathcal{H}_{1}$ and $\mathbf{C}_{\mathbf{y}}|\mathcal{H}_{1}$ as
\begin{equation} \label{Eq4}
    \begin{aligned}
        \mathbf{R}_{\mathbf{y}}|\mathcal{H}_{1}&=\mathbb{E}\Big[\big(\mathbf{H}\mathbf{s}_k+\mathbf{w}_k\big)\big(\mathbf{H}\mathbf{s}_k+\mathbf{w}_k\big)^H\Big]\\
        &=\mathbf{H}\mathbf{R}_{\mathbf{s}}\mathbf{H}^H+\text{diag}\{\sigma^2_{1},\sigma^2_{2},\ldots,\sigma^2_{M}\},\\
        \mathbf{C}_{\mathbf{y}}|\mathcal{H}_{1}&=\mathbb{E}\Big[\big(\mathbf{H}\mathbf{s}_k+\mathbf{w}_k\big)\big(\mathbf{H}\mathbf{s}_k+\mathbf{w}_k\big)^T\Big]\\
        &=\mathbf{H}\mathbf{C}_{\mathbf{s}}\mathbf{H}^T.
    \end{aligned}
\end{equation}
where $\mathbf{R}_{\mathbf{s}}$ and $\mathbf{C}_{\mathbf{s}}$ are respectively the standard covariance and complementary covariance matrices of the transmitted signal $\mathbf{s}_k$. It is clear from the above derivation that $(r_{mm}\in \mathbb{R})>0$, $(r_{mn}=r^{\ast}_{nm})\in \mathbb{C}$ and $(c_{mn}=c_{nm})\in \mathbb{C}$, where $r_{mn}$ and $c_{mn}$ denote the $(m,n)$th element of $\mathbf{R}_{\mathbf{y}}$ and $\mathbf{C}_{\mathbf{y}}$, respectively. Meanwhile, we can notice that under $\mathcal{H}_{0}$, $\mathbf{R}_{\mathbf{y}}|\mathcal{H}_{0}$ is a diagonal matrix and $\mathbf{C}_{\mathbf{y}}|\mathcal{H}_{0}$ is a zero matrix. However, under the alternative hypothesis $\mathcal{H}_{1}$, the off-diagonal elements of $\mathbf{R}_{\mathbf{y}}|\mathcal{H}_{1}$ and all the elements of $\mathbf{C}_{\mathbf{y}}|\mathcal{H}_{1}$ are nonzeros. Combing all above observations, we let  
\begin{equation} \label{Eq5}
    T = \sum_{n-m = 1}^{M-1} \frac{|r_{mn}|^2}{r_{mm}r_{nn}}+\sum_{m = 1}^{M} \frac{|c_{mm}|^2}{2r^2_{mm}}+\sum_{n-m = 1}^{M-1} \frac{|c_{mn}|^2}{r_{mm}r_{nn}},
\end{equation}
If the signal $\mathbf{s}_k$ is absent, $T=0$. While if the signal $\mathbf{s}_k$ is present, $T>0$. Hence, $T$ can be leveraged to distinguish between hypotheses $\mathcal{H}_{0}$ and $\mathcal{H}_{1}$.

It should be noted that, in practical scenarios, the standard covariance and complementary covariance matrices (i.e., $\mathbf{R}_{\mathbf{y}}$ and $\mathbf{C}_{\mathbf{y}}$) can only be computed via a limited number of samples, as given by
\begin{equation} \label{Eq6} 
        \hspace{-3mm}\hat{\mathbf{R}}_{\mathbf{y}}=\frac{1}{K} \sum_{k = 1}^{K}\mathbf{y}_k\mathbf{y}^H_k~\text{and}~
        \hat{\mathbf{C}}_{\mathbf{y}}=\frac{1}{K} \sum_{k = 1}^{K}\mathbf{y}_k\mathbf{y}^T_k.
\end{equation}
Therefore, we propose a NCC method based on \eqref{Eq5} as
\begin{equation} \label{Eq7}
    T_{N} = \sum_{n-m = 1}^{M-1} \frac{|\hat{r}_{mn}|^2}{\hat{r}_{mm}\hat{r}_{nn}}+\sum_{m = 1}^{M} \frac{|\hat{c}_{mm}|^2}{2\hat{r}^2_{mm}}+\sum_{n-m = 1}^{M-1} \frac{|\hat{c}_{mn}|^2}{\hat{r}_{mm}\hat{r}_{nn}},
\end{equation}
where $T_{N}$ is the test statistic of the proposed NCC method, $\hat{r}_{mn}$ and $\hat{c}_{mn}$ denote the elements of $\hat{\mathbf{R}}_{\mathbf{y}}$ and $\hat{\mathbf{C}}_{\mathbf{y}}$ at the $m$th row and the $n$th column, respectively. Given a predefined decision threshold $\lambda_{N}$, if $T_{N} \geq \lambda_{N}$, the signal $\mathbf{s}_k$ is present (``$\mathcal{H}_{1}$'' decision); otherwise, signal $\mathbf{s}_k$ is not present (``$\mathcal{H}_{0}$'' decision).
\subsection{False Alarm Probability and Decision Threshold}
In this subsection, we will derive the asymptotic distribution of the NCC statistic under the null hypothesis, which enable us to calculate the decision threshold for each given false alarm probability.
\begin{lemma}
With a sufficiently large $K$, the random variable $2KT_{N}$ under hypothesis $\mathcal{H}_{0}$ approximately follows the central chi-square distribution with $2M^2$ degrees of freedom (DoFs), i.e.,
\begin{equation}
    2KT_{N}|\mathcal{H}_{0} \sim  \chi_{2M^2}^2,
\end{equation}
where $\chi_{2M^2}^2$ is the central chi-square distribution with $2M^2$ DoFs.
\end{lemma}
\begin{IEEEproof}
Under hypothesis $\mathcal{H}_{0}$, the received signal vector $\mathbf{y}_k$ only consists of noise vector $\mathbf{w}_k$, i,e., $\mathbf{y}_k=\mathbf{w}_k$. Accordingly, $\mathbf{y}_k\sim \mathcal{CN}(\mathbf{0}, \mathbf{R}_{w})$. Denoting the $m$th and $n$th element of the received signal vector $\mathbf{y}_k$ by $y_m(k)$ and $y_n(k)$, respectively. Then $y_m(k)\sim \mathcal{CN}(0, \sigma^2_{m})$ and $y_n(k)\sim \mathcal{CN}(0, \sigma^2_{n})$ are independent for $m\neq n$. In what follows, we divide the proof procedure into the two steps for the convenience of derivations.

1) In the first step, the distributions of the random variables $\{\hat{r}_{mm}\}_{m=1}^{M}$, $\{\hat{r}_{mn}\}_{n-m=1}^{M-1}$, $\{\hat{c}_{mm}\}_{m=1}^{M}$ and $\{\hat{c}_{mn}\}_{n-m=1}^{M-1}$ will be derived. Since $\hat{r}_{mn}$ and $\hat{c}_{mn}$ are respectively the $(m,n)$th element of $\hat{\mathbf{R}}_{\mathbf{y}}$ and $\hat{\mathbf{C}}_{\mathbf{y}}$. Hence, from \eqref{Eq6}, we notice that $\hat{r}_{mn}$ and $\hat{c}_{mn}$ can be respectively expressed as $\hat{r}_{mn}=\frac{1}{K}\sum_{k=1}^{K}y_m(k)y_n^{\ast}(k)$ and $\hat{c}_{mn}=\frac{1}{K}\sum_{k=1}^{K}y_m(k)y_n(k)$. Because $y_m(k)\sim \mathcal{CN}(0, \sigma^2_{m})$ and $y_n(k)\sim \mathcal{CN}(0, \sigma^2_{n})$, therefore, we obtain $\mathbb{E}[y_m(k)y_n(k)]=0$ and
\begin{flalign}
\mathbb{E}[y_m(k)y_n^{\ast}(k)]=\begin{cases}
    \sigma_{m}^2~\text{or}~\sigma_{n}^2, &n = m~\text{or}~m = n;\\
    0, &\text{others}.
    \end{cases}
\end{flalign}
Using the expression of the statistical variance $\mathbb{D}[r]= \mathbb{E}[|r|^2]- |\mathbb{E}[r]|^2$and invoking the following formula \cite{14}:
\begin{align}\label{Eq10}
    \hspace{-2mm}\mathbb{E}[r_{1}r_{2}r_{3}&r_{4}] =\mathbb{E}[r_{1}r_{2}]\mathbb{E}[r_{3}r_{4}] + \mathbb{E}[r_{1}r_{3}]\mathbb{E}[r_{2}r_{4}]  \notag \\
    &+ \mathbb{E}[r_{1}r_{4}]\mathbb{E}[r_{2}r_{3}]-2\mathbb{E}[r_{1}]\mathbb{E}[r_{2}]\mathbb{E}[r_{3}]\mathbb{E}[r_{4}],
\end{align}
we can get
\begin{flalign}
\mathbb{D}[y_{m}(k)y_{n}^{\ast}(k)]=\begin{cases}
    \sigma_m^4~\text{or}~\sigma_n^4, &n = m~\text{or}~m = n;\\
    \sigma_m^2\sigma_n^2, &\text{others}.
    \end{cases}
\end{flalign}
and
\begin{flalign}
\mathbb{D}[y_{m}(k)y_{n}(k)]=\begin{cases}
    2\sigma_m^4~\text{or}~2\sigma_n^4, &n = m~\text{or}~m = n;\\
    \sigma_m^2\sigma_n^2, &\text{others}.
    \end{cases}
\end{flalign}
In \eqref{Eq10}, $r_{1}, r_{2}, r_{3}, r_{4}$ are the complex Gaussian random variables. Therefore, with a sufficiently large $K$, by using the central limit theorem, we can obtain
\begin{flalign}
        &\hat{r}_{mm} \sim \mathcal{N}(\sigma_m^2, \frac{\sigma_m^4}{K}),~\hat{r}_{mn} \sim \mathcal{CN}(0, \frac{\sigma_m^2\sigma_n^2}{K}), \label{Eq13}\\
        &\hat{c}_{mm} \sim \mathcal{CN}(0, \frac{2\sigma_m^4}{K})~\text{and}~\hat{c}_{mn} \sim \mathcal{CN}(0, \frac{\sigma_m^2\sigma_n^2}{K}). \label{Eq14}
\end{flalign}
Additionally, with the aid of \eqref{Eq10}, it is easy to verify that the random variables $\{\hat{r}_{mm}\}_{m=1}^{M}$, $\{\hat{r}_{mn}\}_{n-m=1}^{M-1}$, $\{\hat{c}_{mm}\}_{m=1}^{M}$ and $\{\hat{c}_{mn}\}_{n-m=1}^{M-1}$ are independent of each other.

2) From \eqref{Eq13} and \eqref{Eq14}, we can verify that
\begin{flalign}
        &\hspace{-0.7em}\Big\{\frac{\hat{r}_{mm}}{\sigma^2_m}\Big\}\sim \mathcal{N}(1, \frac{1}{K}),\Big\{\frac{\hat{r}_{mn}}{\sigma_m \sigma_n}, m< n\Big\}\sim \mathcal{CN}(0, \frac{1}{K}), \label{Eq15}\\
        &\hspace{-0.7em}\Big\{\frac{\hat{c}_{mm}}{\sigma^2_m}\Big\}\sim \mathcal{CN}(0, \frac{2}{K}),\Big\{\frac{\hat{c}_{mn}}{\sigma_m \sigma_n}, m< n\Big\}\sim \mathcal{CN}(0, \frac{1}{K}). \label{Eq16}
\end{flalign}
Notice from above that, $\Big\{\frac{\hat{r}_{mm}}{\sigma^2_m}\Big\}\sim \mathcal{N}(1, \frac{1}{K})$, thereby, for a large $K$, we obtain $\hat{r}_{mm}\thickapprox \sigma^2_m$ and $\hat{r}_{nn}\thickapprox \sigma^2_n$. Thus
\begin{equation} \label{Eq17}
\sqrt{\hat{r}_{mm}}\thickapprox \sigma_m~\text{and}~\sqrt{\hat{r}_{nn}}\thickapprox \sigma_n.
\end{equation}
Taking \eqref{Eq17} into \eqref{Eq15} and \eqref{Eq16} yields
\begin{flalign}
        &\Big\{\frac{\hat{r}_{mn}}{\sqrt{\hat{r}_{mm}}\sqrt{\hat{r}_{nn}}}\thickapprox\frac{\hat{r}_{mn}}{\sigma_m \sigma_n}, m< n\Big\}\sim \mathcal{CN}(0, \frac{1}{K}), \\
        &\Big\{\frac{\hat{c}_{mm}}{\hat{r}_{mm}}\thickapprox\frac{\hat{c}_{mm}}{\sigma^2_m}, m=1,2,\ldots,M\Big\}\sim \mathcal{CN}(0, \frac{2}{K}), \\
        &\Big\{\frac{\hat{c}_{mn}}{\sqrt{\hat{r}_{mm}}\sqrt{\hat{r}_{nn}}}\thickapprox\frac{\hat{c}_{mn}}{\sigma_m \sigma_n}, m< n\Big\}\sim \mathcal{CN}(0, \frac{1}{K}).
\end{flalign}
Moreover, we have $\Big\{\frac{2K|\hat{r}_{mn}|^2}{\hat{r}_{mm}\hat{r}_{nn}}=\Big|\frac{\sqrt{2K}\hat{r}_{mn}}{\sqrt{\hat{r}_{mm}}\sqrt{\hat{r}_{nn}}}\Big|^2, m< n\Big\}\sim \chi_{2}^2$, $\Big\{\frac{2K|\hat{c}_{mm}|^2}{2\hat{r}_{mm}^2}=\Big|\frac{\sqrt{2K}\hat{c}_{mm}}{\sqrt{2}\hat{r}_{mm}}\Big|^2, m=1,2,\ldots,M\Big\}\sim \chi_{2}^2$ and $\Big\{\frac{2K|\hat{c}_{mn}|^2}{\hat{r}_{mm}\hat{r}_{nn}}=\Big|\frac{\sqrt{2K}\hat{c}_{mn}}{\sqrt{\hat{r}_{mm}}\sqrt{\hat{r}_{nn}}}\Big|^2, m< n\Big\}\sim \chi_{2}^2$.  Since all the random variables $\{\hat{r}_{mn}\}_{n-m=1}^{M-1}$, $\{\hat{c}_{mm}\}_{m=1}^{M}$ and $\{\hat{c}_{mn}\}_{n-m=1}^{M-1}$ are statistically independent, hence, $\Big\{\frac{2K|\hat{r}_{mn}|^2}{\hat{r}_{mm}\hat{r}_{nn}}, m< n\Big\}$, $\Big\{\frac{2K|\hat{c}_{mm}|^2}{2\hat{r}_{mm}^2}, m=1,2,\ldots,M\Big\}$ and $\Big\{\frac{2K|\hat{c}_{mn}|^2}{\hat{r}_{mm}\hat{r}_{nn}}, m< n\Big\}$ are approximately independent. Note that the variable $\sum_{n-m = 1}^{M-1} \frac{2K|\hat{r}_{mn}|^2}{\hat{r}_{mm}\hat{r}_{nn}}$ is the sum of $\frac{M(M-1)}{2}$ independent $\chi_{2}^2$ random variables, thus we have $\Big\{\sum_{n-m = 1}^{M-1} \frac{2K|\hat{r}_{mn}|^2}{\hat{r}_{mm}\hat{r}_{nn}}\Big\}\sim \chi_{M(M-1)}^{2}$. Similarly, we also can get $\Big\{\sum_{m = 1}^{M} \frac{2K|\hat{c}_{mm}|^2}{2\hat{r}_{mm}^2}\Big\}\sim \chi_{2M}^{2}$ and $\Big\{\sum_{n-m = 1}^{M-1} \frac{2K|\hat{c}_{mn}|^2}{\hat{r}_{mm}\hat{r}_{nn}}\Big\}\sim \chi_{M(M-1)}^{2}$. Based on the above derivations, the variable $2KT_{N} = \sum_{n-m = 1}^{M-1} \frac{2K|\hat{r}_{mn}|^2}{\hat{r}_{mm}\hat{r}_{nn}}+\sum_{m = 1}^{M} \frac{2K|\hat{c}_{mm}|^2}{2\hat{r}^2_{mm}}+\sum_{n-m = 1}^{M-1} \frac{2K|\hat{c}_{mn}|^2}{\hat{r}_{mm}\hat{r}_{nn}}$ follows the chi-squared distribution with $2M^2$ DOFs, i.e., $2KT_{N}\sim \chi_{2M^2}^{2}$. Thus the false alarm probability $P_f$ for the proposed method can be obtained as
\begin{align}
    P_f &= Pr\{T_{N}>\lambda_{N}|\mathcal{H}_{0}\} \notag \\
    &= Pr\Big\{2KT_{N}>2K\lambda_{N}|\mathcal{H}_{0}\Big\} \notag \\
    &= Pr\Big\{\chi_{2M^2}^{2}>2K\lambda_{N}|\mathcal{H}_{0}\Big\}= Q_{\chi_{2M^2}^2}(2K\lambda_N), \notag
\end{align}
where the function $ Q_{\chi_{2M^2}^2}(2K\lambda_{N})$ is the tail probability of the chi-squared distribution with
$2M^2$ DOFs \cite{15}.
\end{IEEEproof}

Hence, with a predefined $P_f$, the corresponding decision threshold $\lambda_N$ can be given as
\begin{equation}
\lambda_N = \frac{Q_{\chi_{2M^2}^2}^{-1}(P_f)}{2K}.
\end{equation}
in which $Q_{\chi_{2M^2}^2}^{-1}(\cdot)$ is the inverse function of $Q_{\chi_{2M^2}^2}(\cdot)$.
\subsection{Computational Complexity} 
In this subsection, the complexity of the proposed NCC method is evaluated in terms of the number of complex multiplications, and is compared to that of two peer methods, i.e., the NC-HDM and NC-LAV methods. The computational complexity of the proposed NCC method is determined by the three parts. The first part is to calculate the upper triangular part of the the matrix $\hat{\mathbf{R}}_{\mathbf{y}}$, which requires $M(M+1)(K+1)/2$ complex multiplications. The second part is to compute the upper triangular part of the the matrix $\hat{\mathbf{C}}_{\mathbf{y}}$, which requires $M(M+1)(K+1)/2$ complex multiplications. The third part is to calculate the test statistic $T_{N}$, which requires $3M^2+M$ complex multiplications. Hence, the total numbers of required complex multiplications for the proposed method is $M^2(K+4)+M(K+2)$, and the corresponding complexity is in the order of $\mathcal{O}(M^2K)$.

As analyzed in \cite{13}, the complexities of the NC-LAV and NC-HDM methods are respectively in the orders of $\mathcal{O}(M^2K)$ and $\mathcal{O}(M^2K)+\mathcal{O}(M^3)$. Therefore, the proposed NCC method enjoys the same complexity order with the NC-LAV method, and both exhibit lower complexity than the NC-HDM method.
\section{Simulation results} 
We now conduct experiments to compare the performance of the proposed NCC method with the following seven methods: CAV \cite{5}, LMPIT\cite{6}, HDM\cite{7}, VD\cite{8}, SFET\cite{9}, NC-HDM\cite{12}, and NC-LAV\cite{13}. At each run, the elements in $\mathbf{H}$ are drawn independent and identically distributed from $\mathcal{CN}(0, 1)$, and the noise powers $\{\sigma^2_{m}\}_{m=1}^{M}$ are independently drawn from the uniform distribution within $[-\alpha, \alpha]$ dB, while the noise parameter is set as $\alpha=1$ in the simulations. Additionally, the $i$th primary signal $s_i(k)(i=1,\ldots,q)$ is assumed to be independent BPSK constellation with different power $\gamma_i(i=1,\ldots,q)$, and the average signal-to-noise ratio (SNR) is defined as 10log$\Big(\frac{\text{tr}(\mathbf{H}\mathbf{R}_{\mathbf{s}}\mathbf{H}^H)}{\text{tr}(\mathbf{R}_n)}\Big)$, each simulation result is averaged over $10^5$ independent Monte Carlo runs.

Figs. 1(a) and (b) plot the detection probability as a function of SNR for the eight methods under the scenario of unequal per-antenna noise variances when $P_{f} = 0.05$, where in Fig. 1(a), the simulation parameters are set as $q=1$, $M=4$ and $K=100$; Meanwhile, in Fig. 1(b), the simulation parameters are set as $q=3$, $M=8$ and $K=200$. From both figures, we can make the following observations. First, one can clearly observe that the proposed method is superior to all other methods in terms of detection probability. For example, in Fig. 1(a), the proposed NCC method achieves at least 0.5 dB gain compared with the other methods at detection probability of 80\%. Moreover, in Fig. 1(b), when SNR $=-11$ dB, the proposed method offers higher detection probability than its counterparts by at least 10\%. Second, it is observed that increasing the SNR can produce a performance gain to all compared methods. Figs. 2(a) and (b) display the receiver operating characteristic (ROC) curves for all compared methods (i.e., detection probability versus false alarm probability), where SNR $=-9$ dB and SNR $=-11$ dB are set in Figs. 2(a) and (b), respectively. There we first see that increasing the false alarm probability is beneficial for all considered methods, i.e., as $P_f$ increases, the detection probabilities of all methods progressively go to one. Next we see that the detection performance of NCC is noticeably better than that of other methods. For example, in both figures, to attain a detection probability of 0.8, the proposed NCC method requires lower $P_f$ than its counterparts.

\begin{figure}[t]
\setlength{\abovecaptionskip}{0.05cm}
\setlength{\belowcaptionskip}{-0.5cm}
\begin{minipage}{4.45cm}
\centering
\centerline{\includegraphics[width=1.88in]{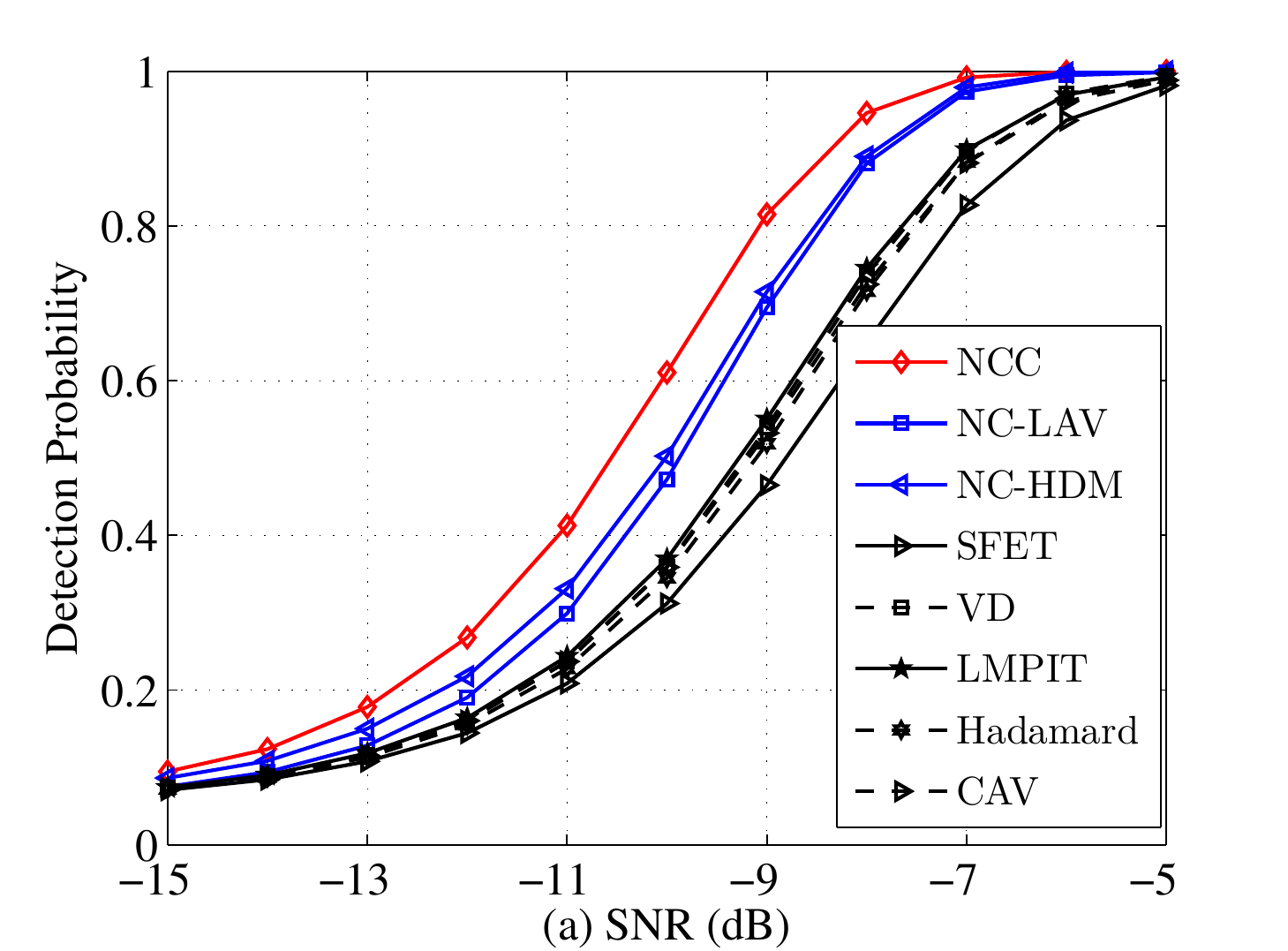}}
\end{minipage}%
\hspace{-0.18cm}
\begin{minipage}{4.45cm}
\centering
\centerline{\includegraphics[width=1.88in]{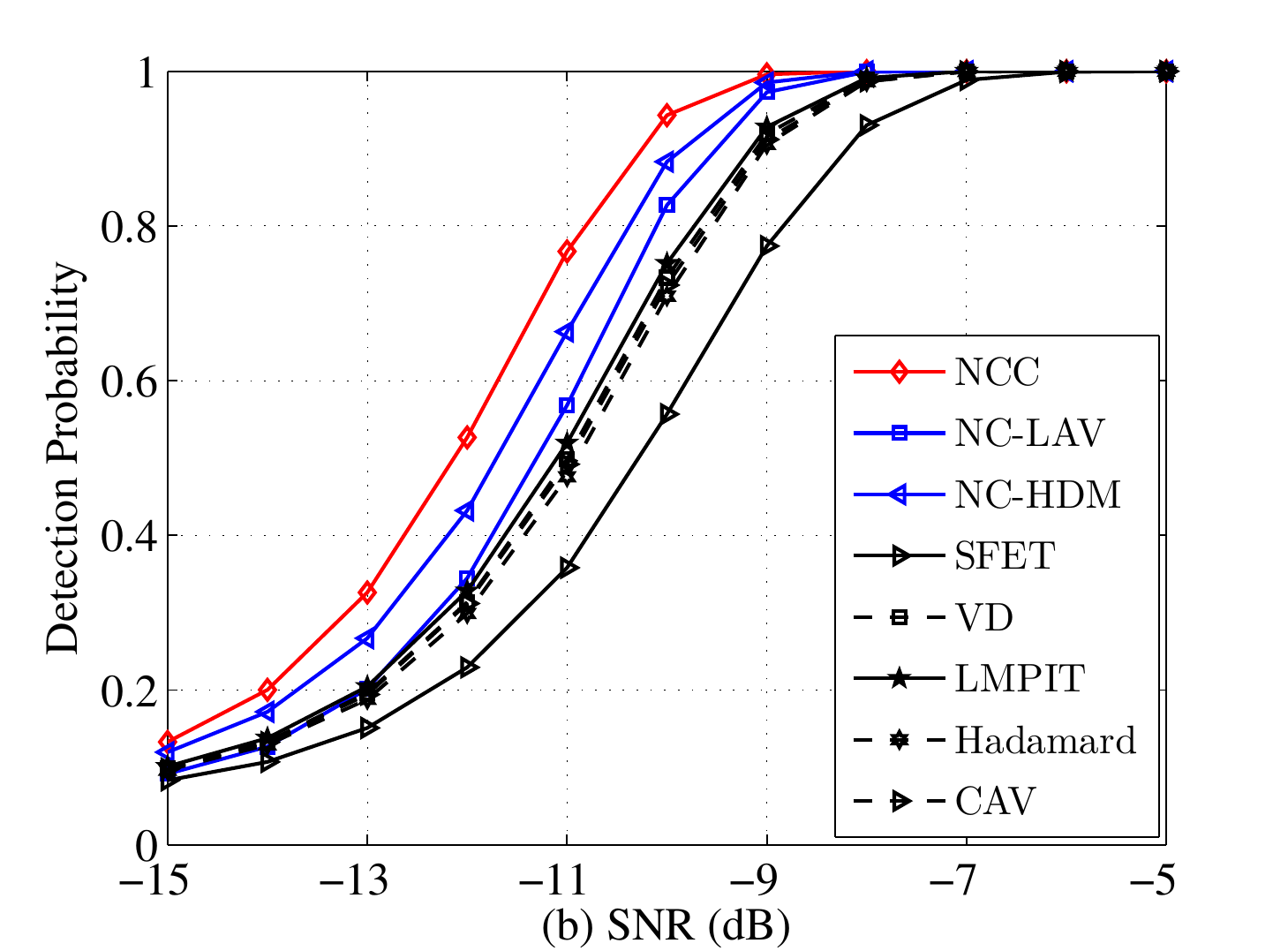}}
\end{minipage}
\caption*{\footnotesize Fig.1. Detection probability versus SNR, where (a) $q=1$, $M=4$ and $K=100$; (b) $M=8$, $K=200$, $q=3$ and $[\gamma_1, \gamma_2, \gamma_3]=[3,1,0]$ dB.} 
\end{figure}

\emph{Discussion on simulation results:} From Figs. 1(a)-1(b) and 2(a)-2(b), we can see that the NCC and NC-HDM have better detection performance than that of other methods, this is because both NCC and NC-HDM methods can exploit the additional information contained in the complementary covariance matrix of received samples, leads to considerable performance gain. Although the NC-LAV method is devised to exploit the NC characteristic of the primary signals. However, this method is developed under the assumption that the noise variances at all antennas are identical, thereby it cannot provide a desirable performance in the scenario of unequal per-antenna noise variances.
\begin{figure}[t]
\setlength{\abovecaptionskip}{0.05cm}
\setlength{\belowcaptionskip}{-0.5cm}
\begin{minipage}{4.45cm}
\centering
\centerline{\includegraphics[width=1.88in]{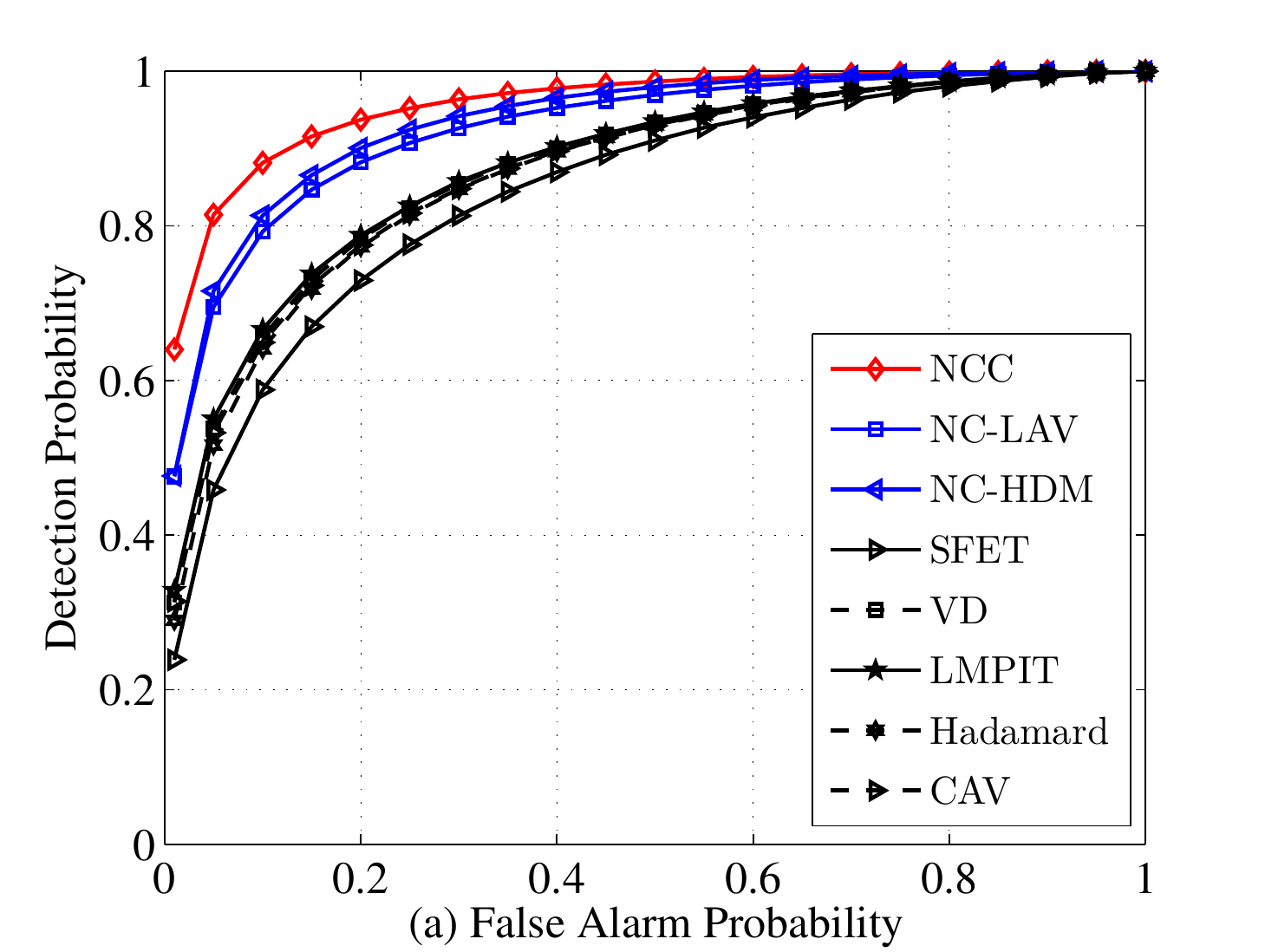}}
\end{minipage}%
\hspace{-0.18cm}
\begin{minipage}{4.45cm}
\centering
\centerline{\includegraphics[width=1.88in]{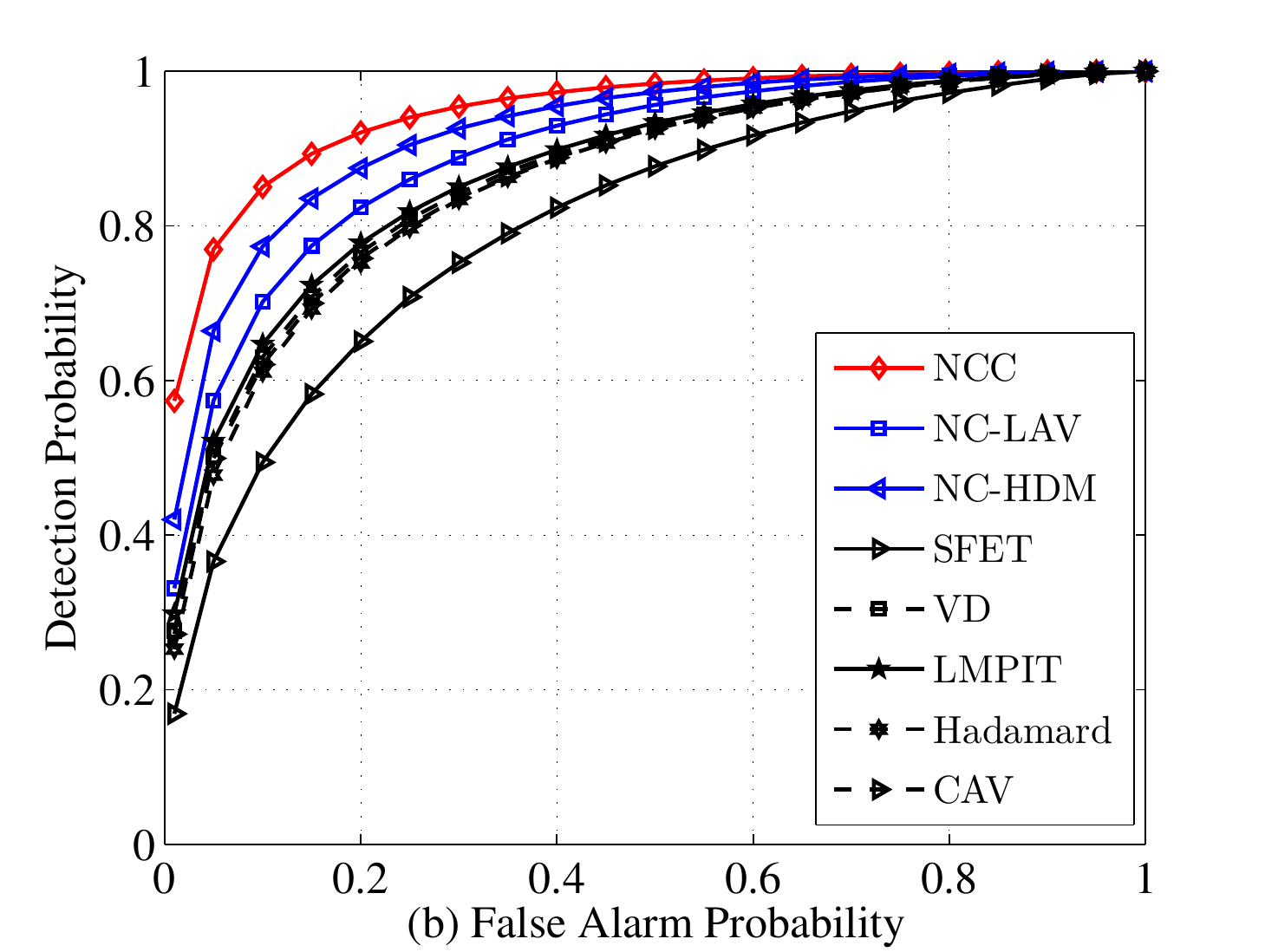}}
\end{minipage}
\caption*{\footnotesize Fig.2. ROC curves for all compared methods, where (a) $q=1$, $M=4$ and $K=100$; (b) $M=8$, $K=200$, $q=3$ and $[\gamma_1, \gamma_2, \gamma_3]=[3,1,0]$ dB.} 
\end{figure}
\section{Conclusion}
In this letter, the spectrum sensing problem in multiantenna CR networks with NC primary signals was addressed under the scenario of unequal per-antenna noise variances, and a low-complexity NCC method was proposed to exploit the NC characteristic of the primary signals. Numerical results show that the NCC method is superior to the state-of-the-art detection methods.

\renewcommand\refname{References}
\footnotesize
\bibliographystyle{ieeetran}
\bibliography{reference} 
\end{document}